\documentclass[%
 reprint,
 amsmath,amssymb,
 aps,
]{revtex4-2}
\usepackage{graphicx}
\usepackage{color}
\usepackage{dcolumn}
\usepackage{latexsym}
\usepackage[normalem]{ulem}
\usepackage{hyperref,amssymb}
\usepackage{url}
\usepackage{booktabs}
\usepackage{color,soul}
\usepackage{graphicx}
\usepackage{siunitx}
\usepackage{verbatim}
\usepackage{multirow}
\usepackage{amsmath}
\newcommand{\beq}{\begin{eqnarray}}
\newcommand{\eeq}{\end{eqnarray}}
\usepackage{mathrsfs}
\usepackage{float,soul}
\usepackage[dvipsnames]{xcolor}
\usepackage{orcidlink}
\usepackage{mathtools}
\usepackage{slashed}
\usepackage{physics}	
\usepackage{graphicx}   
\usepackage{epstopdf}
\usepackage{subfigure}  
\usepackage{hyperref}   
\usepackage{bbold}
\usepackage{wasysym}
\usepackage{feynmp}
\usepackage{hyperref}
\hypersetup{colorlinks,
}

\usepackage{siunitx}
\usepackage{color,xcolor}
\usepackage{soul}
\usepackage{color,xcolor}
\usepackage{relsize}
\soulregister{\cite}7
\soulregister{\ref}7
\soulregister{\eqref}7
\usepackage{amsmath}

\usepackage{booktabs}

\DeclareSIUnit\angstrom{\text{\AA}}

\begin{document}

\title{\Large 
Physical signatures of supercritical fluid boundaries}

\author{Sha Jin\orcidlink{0000-0003-3398-9727}$^{1,2,\parallel}$}
\email{kingsian@sjtu.edu.cn}
\author{Xinyang Li\orcidlink{0000-0002-0226-0581}$^{3,4,\parallel}$}
\email{lixinyang@itp.ac.cn}
\author{Xue Fan\orcidlink{0000-0001-5943-6042}$^{5,6}$}
\email{fanxue2015@shu.edu.cn}
\author{Matteo Baggioli\orcidlink{0000-0001-9392-7507}$^{1,2}$}
\email{b.matteo@sjtu.edu.cn}
\author{Yuliang Jin\orcidlink{0000-0001-8310-4582}$^{4,7,8},$}
\email{yuliangjin@mail.itp.ac.cn}

\altaffiliation[$\parallel$]{Contributed equally to this work}

\address{$^1$Wilczek Quantum Center, Shanghai Jiao Tong University, Shanghai 200240, China}
\address{$^2$Shanghai Research Center for Quantum Sciences, Shanghai 201315, China}
\address{$^3$Peng Huanwu Collaborative Center for Research and Education, Beihang University, Beijing 100191, China}
\address{$^4$Institute of Theoretical Physics,
 Chinese Academy of Sciences, Beijing 100190, China}
\address{$^5$College of Materials, Shanghai Dianji University, Shanghai 201306, China}
\address{$^6$Materials Genome Institute, Shanghai University, Shanghai 200444, China}
\address{$^7$School of Physical Sciences, University of Chinese Academy of Sciences, Beijing 100049, China}
\address{$^8$Wenzhou Institute, University of Chinese Academy of Sciences, Wenzhou, Zhejiang 325000, China}

\date{\today}
\begin{abstract}
In the supercritical fluid (SCF) region, at temperatures and pressures above the critical point, the thermodynamic singularity separating liquids and gases no longer exists. Recent arguments based on thermodynamics and critical scalings  have revived the proposal that the SCF constitutes an intermediate state of matter, separated from the liquid and gas by two supercritical boundaries, the $L^\pm$ lines. However, until now, the nature of the supercritical state and the physical signatures of these boundaries have remained elusive. Here, we demonstrate that the SCF is characterized by distinct structural, transport, and dynamical behavior. Specifically, the spatial arrangement of particles—captured by the radial distribution function—as well as the diffusion coefficient, shear viscosity, and velocity autocorrelation function in the SCF regime are qualitatively different from those in both the liquid and gas states and exhibit physical signatures upon crossing the $L^\pm$ lines. Our theoretical predictions are supported
by molecular dynamics simulations of argon and are further supported by existing experimental evidence. These results provide a clear physical foundation for a refined phase diagram of matter in the supercritical region, comprising three distinct states—gas, supercritical fluid, and liquid—separated by two crossover boundaries obeying universal scaling laws.
\end{abstract}

\maketitle

\section{\label{sec:Introduction}Introduction}
Liquids and gases are arguably among the most common phases of matter \cite{tabor1991gases} and have been studied and probed experimentally for centuries. At low-enough pressure and temperature, they are separated by a coexistence line defined by a thermodynamic first-order phase transition \cite{domb2000phase}. Nevertheless, the coexistence line has finite extension and ends at the so-called critical point (e.g., $T_c\approx 647$ K and $P_c\approx 220$ bar for water), above which lies a large part of the phase diagram---the supercritical fluid (SCF) region \cite{kiran2012supercritical,proctor2020liquid}---which was first discovered by Baron de la Tour in 1822 \cite{de1822expose}. In this SCF region, liquid and gas phases can be connected without encountering any thermodynamic singularity, which raises the long-standing question of what really separates a liquid from a gas \cite{10.1063/PT.3.1796}, if anything, and whether a supercritical fluid constitutes a distinct phase of matter.

A commonly considered scenario is that the SCF region contains liquid-like and gas-like states separated in the phase diagram by a single separation line. Importantly, this line does not correspond to any thermodynamic singularity and therefore represents a continuous crossover, making both its definition and precise location inherently subtle. Indeed, multiple distinct and noncoincident definitions of this crossover line have been proposed in the literature, including the Widom line, Frenkel line, and Fisher-Widom line, among others~\cite{xu2005relation,jones1956specific,luo2014behavior,banuti2017similarity,corradini2015widom,gallo2014widom,fisher1969decay,vega1995location,evans1993asymptotic,PhysRevE.85.031203,brazhkin2013liquid,bolmatov2013thermodynamic,bolmatov2014structural,fomin2018dynamics,nishikawa1995correlation,matsugami2014theoretical,ploetz2019gas,woodcock2013observations,woodcock2016thermodynamics,brazhkin2012supercritical,huang2023revealing,ouyang2024complex}.

\begin{figure}[t]
    \centering
\includegraphics[width=\linewidth]{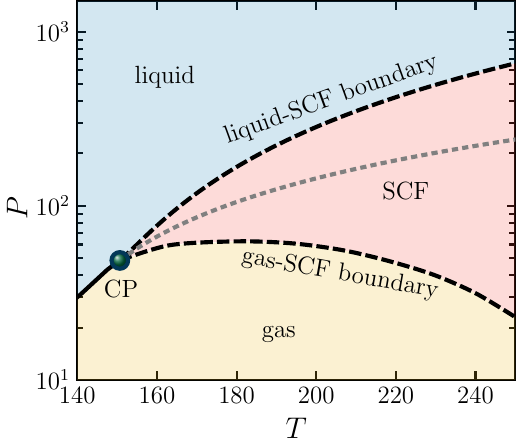}
    \caption{\textbf{Phase diagram of argon.} The critical point (CP) is located at $(T_c,P_c)\approx (151\,\text{K}, 48.55\,\text{bar})$ \cite{1967Critical}.  The liquid-gas coexistence line (solid black line) is obtained from the NIST database~\cite{NIST}. 
     The gray dotted line is the critical isochore, which represents the supercritical continuation of the coexistence line. 
    The two black dashed lines are $L^{\pm}$ lines, which are computed using the NIST data and the thermodynamic criterion proposed in \cite{li2024thermodynamic}, and represent the boundaries between the liquid, gas, and SCF phases.
    }
    \label{fig:1}
\end{figure}

An alternative scenario, summarized in Fig.~\ref{fig:1}, posits that the liquid and gas phases are separated by a finite crossover region, bounded by two boundary lines~\cite{banuti2015crossing,maxim2021thermodynamics,wang2021three,li2024thermodynamic}, thereby rendering the SCF in this region distinguishable from both a liquid and a gas. A two-boundary picture has been explored, for instance, in the Widom delta scenario \cite{Ha2018,10.1063/1.5086467,Ha2020}, where the crossover lines are associated with the percolation loci of gas-like and liquid-like clusters. In this framework, however, the intermediate SCF region is not treated as a distinct state of matter but rather as a gas–liquid coexistence regime. Moreover, the definition of such clusters is often ambiguous and typically relies on machine-learning-based classifications \cite{Ha2018}, lacking a definitive interpretation of the associated physical signatures (see, however, \cite{ASHARCHUK202549}).

On the other hand, more recent studies have established the thermodynamic origin of these two supercritical boundary lines ($L^\pm$ lines) ~\cite{li2024thermodynamic} and their universal scalings in liquid-gas phase transitions~\cite{li2024thermodynamic}, Ising models~\cite{li2024thermodynamic}, quantum phase transitions~\cite{wang2024quantum, lv2024quantum}, antiferromagnet materials~\cite{Liu2026Ising}, charged black holes~\cite{wang2025analogous}, and liquid-liquid phase transitions~\cite{wang2025analogous}. A recent comprehensive study~\cite{torres2026comprehensive} benchmarked the $L^{\pm}$ lines against extensive experimental and simulation data across 15 fluids of diverse molecular types, demonstrating that the $L^{+}$ and $L^{-}$ lines consistently describe the liquid-SCF-gas crossover in the supercritical region. 

If, however, the SCF constitutes an independent state of matter, as implied by the second scenario~\cite{li2024thermodynamic}, then a natural question arises: \textit{what defines this state, and what physical signatures mark the crossing of the two supercritical boundaries}? Here, we address this question by proposing a specific two-boundary scenario and examining whether it is consistent with simulations and with existing experimental data. We note that these two crossover boundaries are not sharp, and we do not systematically investigate whether the same simulation results could be alternatively interpreted by a single separation line scenario. Nevertheless, we show that the SCF exhibits unique structural, transport, and dynamical properties, fundamentally different from those of both liquids and gases. These differences produce clear, albeit noncritical, physical signatures when crossing each of the two supercritical boundaries, consistently observed in both simulations and experiments.

\section{\label{sec:Phase diagram}Phase diagram and supercritical boundaries}
\begin{figure}[t]
    \centering
\includegraphics[width=0.9\linewidth]{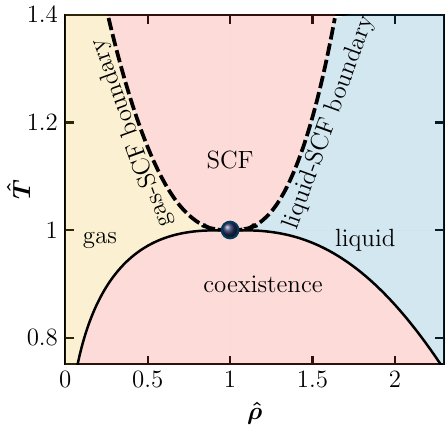}
    \caption{\textbf{$\hat T-\hat \rho$ Phase diagram of argon.} The solid line represents the gas-liquid coexistence curve, and the dashed lines denote the supercritical crossover boundaries (gas-SCF boundary and liquid-SCF boundary). Here, $\hat T = T/T_c$ and $\hat{\rho} = \rho/\rho_c$.
    }
    \label{fig:T-rho}
\end{figure}

We consider Ar as a prototypical simple system and perform extensive molecular dynamics (MD) simulations.
In Fig.~\ref{fig:1}, we present the pressure-temperature phase diagram of Ar (see Fig.~\ref{fig:T-rho} for the corresponding temperature-density phase diagram). The critical point is located at $(T_c,P_c)\approx (151\,\text{K}, 48.55\,\text{bar})$ \cite{1967Critical}.
The liquid-gas coexistence line (the black line in Fig.~\ref{fig:1}) continues into the supercritical regime (the gray dotted line), which typically shows a high symmetry revealed by maxima of density fluctuations under constant temperature conditions~\cite{xu2005relation, nishikawa1995correlation, ploetz2019gas}. In this study, we regard all supercritical lines defined in this spirit as the continuation of the coexistence line, including the critical isochore, Widom line~\cite{xu2005relation}, Nishikawa line~\cite{nishikawa1995correlation}, and symmetry line~\cite{ploetz2019gas}. These lines locate slightly differently in the phase diagram (see \cite{li2024thermodynamic}), but all around the middle of the SCF phase. Importantly, in our framework, and different from previous interpretation, those are not considered as the ``boundary line'' between liquid-like and gas-like states.

As shown in Fig.~\ref{fig:1}, we propose that the gas, SCF, and liquid phases are separated by two boundary lines (black dashed lines). (i) The liquid-SCF boundary roughly corresponds to the Fisher-Widom line~\cite{vega1995location}, defined based on the behavior of the radial distribution function $g(r)$ (a structural viewpoint), or the Frenkel line defined based on dynamical viewpoints. The location of the Frenkel line depends on the specific criterion employed: the Frenkel-1 line is defined by the equality $\tau_0 = \tau$ between the vibration time $\tau_0$ and the liquid relaxation time $\tau$~\cite{PhysRevE.85.031203}, the Frenkel-2 line corresponds to the disappearance of oscillations and minima of the velocity autocorrelation function~\cite{brazhkin2013liquid}, and the Frenkel-3 line is derived based on isochoric heat capacity  
$C_V$ (for monatomic systems such as argon, the criterion is $C_V= 2k_{\rm B}$)~\cite{PhysRevE.85.031203}. (ii) The gas-SCF boundary, which is the focus of the present study, has never been individually investigated in previous works. (iii) Several existing considerations have proposed two boundaries, including the $L^{\pm}$ lines~\cite{li2024thermodynamic},  percolation lines~\cite{woodcock2013observations, woodcock2016thermodynamics}, the two boundaries of the pseudoboiling transitional region~\cite{banuti2015crossing, wang2021three}, and the Widom delta~\cite{Ha2018,10.1063/1.5086467,Ha2020}.
Different criteria give nearby, but quantitatively different boundary lines (see \cite{li2024thermodynamic}). 
Again, the reason is that these lines do not correspond to any thermodynamic or dynamic singularities, which means that they cannot be uniquely defined in a mathematically rigorous way even in the thermodynamic limit.  
An exception is the $L^\pm$ lines in quantum phase transitions, where the thermodynamic supercritical crossovers coincide with dynamic singularities under quantum quenches~\cite{wang2024quantum}. 

Importantly, among the above proposals, only the $L^\pm$ lines  display universal scalings in various phase transitions~\cite{li2024thermodynamic, wang2024quantum, lv2024quantum, Liu2026Ising,wang2025analogous}: the scalings of the external field and order parameter along $L^\pm$ lines satisfy $\delta P^{\pm} \sim (T-T_{\rm c})^{\beta + \gamma}$ and  $\delta \rho^{\pm} \sim (T-T_{\rm c})^{\beta}$, where $\beta$ and $\gamma$ are standard critical exponents. For these reasons, below, we adopt the $L^\pm$ lines, introduced in Ref.~\cite{li2024thermodynamic} and shown for the case of argon as black dashed curves in Fig.~\ref{fig:1}. To define the $L^{\pm}$ lines~\cite{li2024thermodynamic}, one first chooses the critical isochore, which is an extension of the coexistence line to the supercritical region, as the reference line. The compressibility $\kappa_T \equiv \frac{P_c}{\rho^{\rm m}_c}\left( \frac{\partial \rho^{\rm m}}{\partial P} \right)_T$ is evaluated along each path parallel to the critical isochore, where $P_c$ and $\rho^{\rm m}_c$ are the critical pressure and critical mass density, respectively. Since $\kappa_T$ is a function of distance $\delta P(\rho^{\rm m}, T) = P(\rho^{\rm m},T) - P(\rho^{\rm m}_c,T)$ and $T$, one can find a temperature $T_{\rm max}(\delta P)$ that maximizes $\kappa_T$ along each path with a fixed $\delta P$. All the $T_{\rm max}(\delta P)$ points under different $\delta P$ together consist of  $L^{\pm}$ lines. To determine $L^\pm$ lines, one needs to know the EOS $P(\rho^{\rm m}, T)$. In this study, we take the EOS of argon from the NIST database~\cite{NIST}. Note that the \(L^\pm\) lines can be determined only in the temperature range where the compressibility shows a peak. Consequently, the applicability of the phase diagram (Fig.~\ref{fig:1}) is restricted to the near-critical region.

To complement the pressure-temperature (\(P\)-\(T\)) phase diagram shown in Fig.~\ref{fig:1}, we present in Fig.~\ref{fig:T-rho} the corresponding density-temperature ($\hat T-\hat \rho$) phase diagram of argon. Quantities are given in reduced parameters ($\hat{T}=T/T_{\rm c}$, $\hat{\rho}=\rho/\rho_{\rm c}$).
This \(\hat T\)-\(\hat \rho\) representation reveals a striking mirror symmetry between the supercritical crossover region (bounded by \(L^-\) and \(L^+\)) and the subcritical liquid-gas coexistence dome~\cite{Li2026Supercritical}. In both cases, the two boundaries merge at the critical point and diverge as one moves away from it. However, the physical nature of the two regions is fundamentally different: in the subcritical domain, the coexistence dome encloses a two-phase region where liquid and gas coexist as distinct thermodynamic phases; in the supercritical region, the SCF is not a two-phase coexistence state because liquid and gas states are indistinguishable.

\section{\label{sec:MD methods} Simulation methods}
For the detailed analyses below, we select two isothermal lines corresponding to $T=195$ K and $T=220$ K, respectively, and consider a large range of pressures that scan from largely above the liquid-SCF boundary to well below the gas-SCF boundary in MD simulations. This will allow us to study in detail the structural, transport, and dynamical properties across the boundaries, revealing their fundamental origin and nature.

We employ the Lennard-Jones (LJ) potential for argon in the LAMMPS package~\cite{THOMPSON2022108171}, which benefits from efficient parallelization for large atom counts. The expression for the LJ potential is given as
\beq
V(r) = 4 \epsilon \left [\left(\frac{\sigma}{r}\right)^{12} - \left(\frac{\sigma}{r}\right)^{6} \right],
\label{eq:V_LJ}
\eeq
where
$\epsilon/k_{\rm B} = 119.5$ K, $\sigma  = \SI{3.405}{\angstrom}$, and $k_{\rm B} = \SI{1.38e-23}{\joule\per\kelvin}$ is the Boltzmann constant. These parameters are the standard ones for argon, and therefore the subsequent simulations are specifically performed for this system. To address potential finite-size effects, we have evaluated systems of 500, 4000, and 32\,000 atoms. Our analysis confirms that 4000 atoms provides a balance between statistical accuracy and computational efficiency for calculating transport properties; consequently, this system size is adopted for the production runs. The initial system is a face-centered cubic lattice of 4000 atoms with a lattice constant of $\SI{5.26}{\angstrom}$. Thirty-five different pressures are simulated from $\SI{5}{\bar}$ to $\SI{1300}{\bar}$. For each target pressure, the system is initially relaxed at $\SI{195}{\K}$ for $\SI{5}{\ns}$ in an isothermal-isobaric (NPT) ensemble to reach thermal equilibrium. Following the NPT stage, the system undergoes an intermediate 500 ps relaxation in the microcanonical (NVE) ensemble at each specific state point to ensure total energy stability and facilitate unconstrained dynamical relaxation. Finally, a canonical (NVT) relaxation is performed to stabilize the system prior to the $\SI{20}{\ns} $ production phase---all measurements are performed in this time period.
The temperature and pressure are regulated via the Nos\'{e}-Hoover methods~\cite{10.1063/1.447334,PhysRevA.31.1695}, with Newton's equations of motion solved numerically using a time step of $\SI{1}{\fs}$. All samples are simulated with periodic boundary conditions. The pressure unit is bar, length unit Angstroms, time unit fs, and energy unit kcal/mol.

\begin{figure*}[t]
    \centering
    \includegraphics[width=\linewidth=0.9]{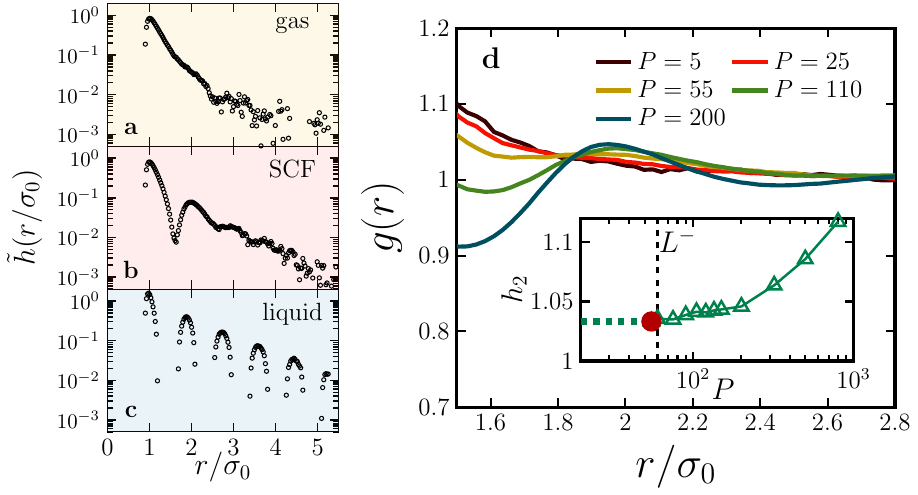}
    \caption{\textbf{Radial distribution functions (RDFs).} \textbf{(a-c)} Typical reduced RDF, $\tilde{h}(r) = r[g(r)-1]$, of gas, SCF, and liquid states measured in simulations of argon (gas: $T=195$ K, $P=20~\text {bar}$; SCF: $T=195$ K, $P=90 ~\text{bar}$; liquid: $T=100$ K, $P=40 ~\text{bar}$). 
The distance $r$ is expressed in units of $\sigma_0$, the position of the first peak. 
    \textbf{(d)} Radial distribution function $g(r)$ at different pressures, zoomed in around the second peak, for $T=195\ \text{K}$. \textbf{(inset)} Height of the second peak $h_2$ as a function of $P$; below $P_2 \approx 55$ bar (red circle), the second peak can not be identified anymore.
    } 
    \label{fig:length}
\end{figure*}

\begin{figure}[h]
    \centering
\includegraphics[width=0.9\linewidth]{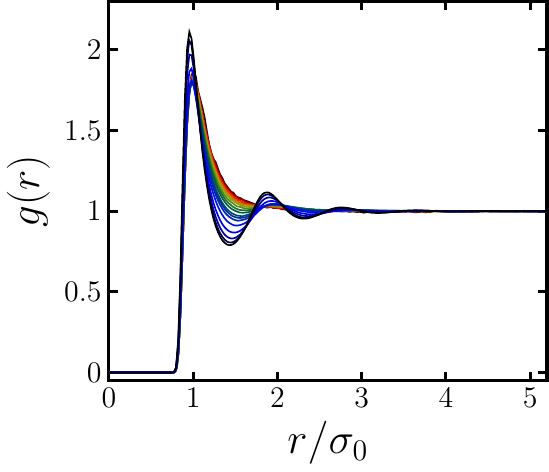}
    \caption{\textbf{Radial distribution functions at  $T=195\ \text{K}$.} Radial distribution functions $g(r)$ for argon are obtained from MD simulations at $P=$\ 5,15, 25, 35, 45, 55, 60, 75, 90, 105,120,135,150, 200,320,500,700,800\ \text{bar} (from red to blue) and $T=195\ \text{K}$.}
   
    \label{fig:gr}
\end{figure}

\section{\label{sec:Structural evidence}Structural evidence: radial distribution functions}

\subsection{Typical radial distribution functions in gas, SCF, and liquid states}

In Figs.~\ref{fig:length}(a)--\ref{fig:length}(c), we present the reduced radial distribution function $\tilde{h}(r) = r[g(r)-1]$, in three distinct states: gas, SCF, and liquid. The gas state exhibits a rapid decay 
of the tail of $\tilde{h}(r)$ without oscillations,
indicating the weak intermolecular correlations. In contrast, the liquid displays exponentially damped (yet persistent) oscillations with multiple distinct peaks, signaling robust short-range order. The SCF exhibits an intermediate structure: a few peaks emerge, yet higher-order coordination shells are absent, indicating loss of order at large distances. This represents a unique structural signature of the SCF, setting it apart from the gas and liquid states.

Based on the above evidence, we propose that the structure of SCF is characterized by sub-short-range (SSR) order. The SSR order is reflected in the radial distribution function $g(r)$ through damped oscillations that are truncated, i.e., the oscillatory behavior of $g(r)$ extends only up to a finite SSR length scale. This viewpoint on structure and, as we will see, also the behavior of  transport and dynamical properties, implies that the SCF can be interpreted as an independent state of matter in a loose sense. This definition is not rigorous because there are no bona fide phase transitions between SCF and gases/liquids and all lines should rather be regarded as smooth crossovers. Nevertheless, these three phases can still be separated by two supercritical boundary lines as discussed in the following.

\subsection{Gas-SCF boundary: disappearance of the second peak}

The gas-SCF boundary corresponds to the locus where the oscillatory behavior in $g(r)$, aside from the first peak, which persists except in the ideal gas limit, disappears. In practice, it can be located by the disappearance of the second peak in $g(r)$, or equivalently the first minimum, as shown by the transition from Fig.~\ref{fig:length}(a) to Fig.\ref{fig:length}(b). 

More specifically, we measure $g(r)$ using MD simulations at $T=195$ K, in the pressure range from 5 bar to 900 bar (see Fig.~\ref{fig:gr}), and we focus on the second peak in $g(r)$ (see Fig.~\ref{fig:length}d). 

The data of $g(r)$  are calculated by averaging over a large number of frames to reduce statistical noise. We extract 500 ps of trajectories with 1 fs intervals, resulting in 500\,000 frames for averaging. These extensive averages ensure that the data are well-converged.

To quantitatively track the evolution of structural features in the radial distribution function $g(r)$, its second peak is fitted using a Gaussian function. Specifically, we employ a Gaussian function of the following form to fit the second peak:
\begin{equation}
f(r)=c +h_2 \exp\!\left[-\frac{(r-r_0)^2}{2\sigma^2}\right],
\end{equation}
with fitting parameters $c, h_2, r_0$ and $\sigma$. Here, $h_2$ and $r_0$ provide the height and position, respectively. When the coefficient $R^2$ of determination is below $0.8$, the Gaussian model no longer provides an adequate description of the feature, indicating  that the second peak cannot be  reliably identified.

Based on the data of $g(r)$, a separation pressure $P_2 \approx 55$ bar is determined: when $P>P_2$, the height of the second peak $h_2(P)$ increases with $P$; when $P<P_2$, the second peak cannot be identified within the given accuracy of the numerical data.

\begin{figure}[t]
    \centering
    \includegraphics[width=\linewidth]{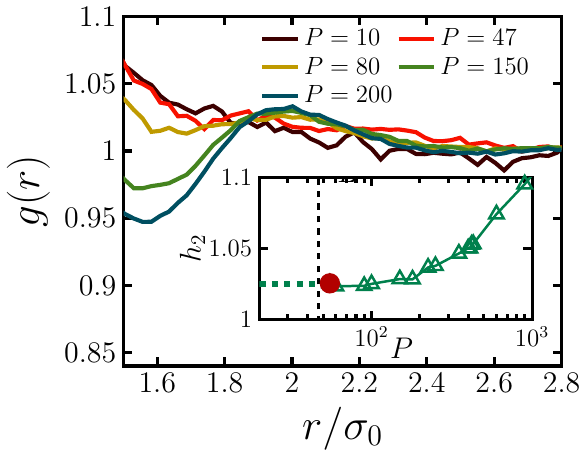}
    \caption{\textbf{Enlarged radial distribution functions at $T=220\ \text{K}$.} Radial distribution function $g(r)$ at different pressures, zoomed in around the second peak, for $T=220\ \text{K}$.
    } 
 
    \label{fig:220gr}
\end{figure}

Below $P_2$, the $g(r)$ has only one peak, $g(r) \approx \exp[-V(r)/k_{\rm B}T]$, where $V(r)$ is the pair interparticle interaction, as given by the low-density theory~\cite{chandler1987introduction}. In this case, $g(r)$ decays beyond the first peak without any oscillations---this is the definition of the gas state. On the other hand, $P_2$ should correspond to the gas-SCF boundary at $T=195$~K. Interestingly,  $(P_2\approx 55 \, {\rm bar}$, $T=195\,~\rm{K})$ coincides with the point $(P^{-} \approx 60 \, {\rm bar}$, $T=195 \, \rm{K})$ on the $L^-$ line estimated independently in ~\cite{li2024thermodynamic} using thermodynamics, confirming $L^-$ as the gas-SCF boundary. We also verify the universality of these findings by considering a second isothermal $T=220 \rm{K}$ cut crossing the $L^\pm$ lines (see Fig.~\ref{fig:220gr}). The second peak emerges around $P_2 = 55 $ bar, consistent with the pressure $P^-= 47$ bar at the $L^-$ line. This observation suggests that, from a structural perspective, the $L^-$ line corresponds to the disappearance of the second peak in $g(r)$.

In the following sections, we reveal that the supercritical boundaries are also reflected in transport properties, specifically, the viscosity $\eta$ and the self-diffusion coefficient $D$, as well as in the dynamical behavior, such as the velocity autocorrelation function (VACF), $Z(t)$. These observations can also be derived from a liquid-state theory based on the memory-function approach and are supported by simulation results. The numerically estimated gas-SCF boundary based on these criteria aligns closely with the lower $L^-$ line independently identified in Ref.~\cite{li2024thermodynamic} using a thermodynamic criterion, suggesting a clear physical significance for the latter.\\

\subsection{Liquid-SCF boundary: crossover from monotonic to exponentially damped oscillatory decay}

From a structural perspective, the Fisher-Widom line \cite{fisher1969decay} corresponds to the liquid-SCF boundary. In liquids, the oscillations in $g(r)$, $\sim\cos(qr)$, although damped, remain up to $r \to \infty$. The liquid-SCF boundary has been alternatively estimated based on dynamic criteria (Frenkel line~\cite{PhysRevE.85.031203}), such as the disappearance of the short-time vibrational motion typical of liquids. The Fisher-Widom line and the Frenkel line are also close to the upper $L^+$ line defined from thermodynamic and scaling considerations~\cite{li2024thermodynamic}. In general, the liquid-SCF boundary corresponds to the boundary beyond which the state loses liquid characteristics, although its precise location depends on the detailed criteria. The small quantitative inconsistency is possibly inevitable (also for the gas-SCF boundary), due to the absence of thermodynamic or dynamic singularities. Nevertheless, these lines are close to each other and can be unambiguously distinguished from the gas-SCF boundary (as we will show in this study). Since the liquid-SCF transition has been extensively studied in the literature, including from the perspective of structural criteria such as the Fisher-Widom line~\cite{fisher1969decay, de1994decay, vega1995location, dijkstra2000simulation, evans1993asymptotic, tarazona2003fisher, stopper2019decay}, we do not focus on it here. Instead, our study concentrates on the gas-SCF boundary, which remains largely unexplored.\\

\begin{figure*}[t]
    \centering
    \includegraphics[width=\linewidth=1]{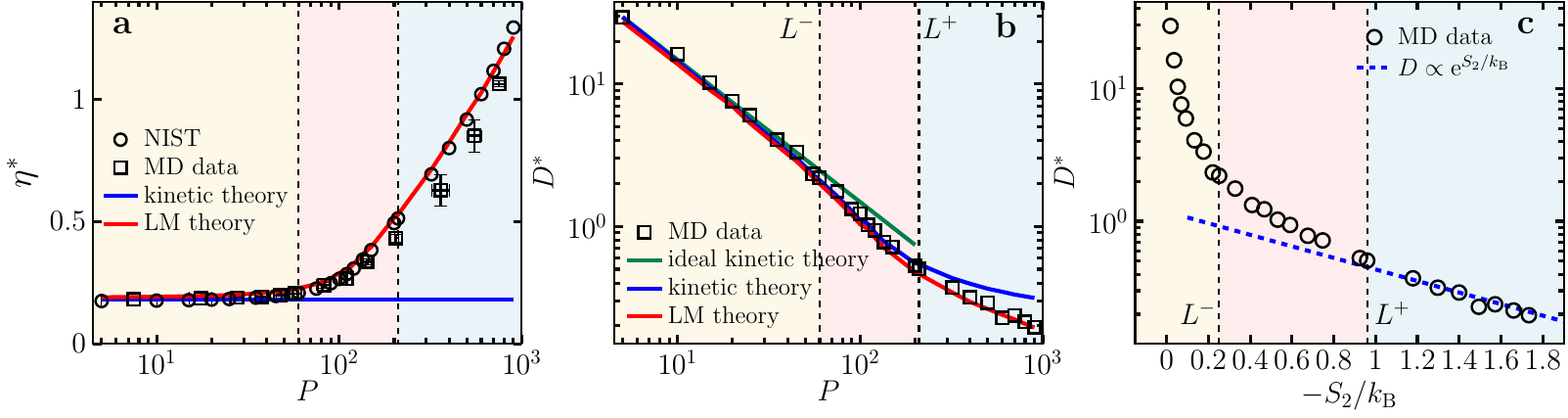}
    \caption{\textbf{Transport coefficients.}
    \textbf{(a)} Reduced viscosity $\eta^*=\eta\left(\sigma^2 / \sqrt{\epsilon m}\right)$,  where $\sigma$ and $\epsilon$ are length and energy parameters in the  Lennard-Jones potential (see Simulation Methods), and $m$ is the atomic mass. 
    The error bars represent the 
    standard error of the mean.
    \textbf{(b)} Reduced diffusion coefficient $D^*=D \sqrt{\left(\frac{m}{\epsilon \sigma^2}\right)}$. In the kinetic theory expression, $D_{0} = \frac{3}{8\rho \sigma^2} \left(\frac{ k_{\rm B}T}{\pi m} \right)^{1/2}$, we have used both the ideal gas form $\rho_{\rm ideal} (P,T)= k_{\rm B} P/T$  (\textit{ideal kinetic theory}), and the exact $\rho(P,T)$ from simulation data (\textit{kinetic theory}). \textbf{(c)} Reduced diffusion coefficient as a function of the two-particle structural entropy $S_2$. The blue dashed line represents an exponential fit of the data above the $L^+$ line, $D = 1.18 e^{S_2/k_{\rm B}}$.}
    \label{fig:eta}
\end{figure*}

\section{\label{sec:Kinetic features} Kinetic features\text{:} transport coefficients}

\subsection{Gas-SCF boundary: deviation from kinetic theory}

Next, we present  signatures of the  $L^\pm$ crossover lines in the transport coefficients. In the dilute limit, transport processes are well described by the well-known kinetic theory~\cite{chapman1990mathematical, hansen2013theory}. However, beyond the gas-SCF boundary, non-negligible corrections to the transport coefficients emerge---in particular, to the viscosity $\eta$, where the excess part $\eta_{\rm ex}$ that depends on the interaction potential and configurational structure becomes dominant compared to the kinetic part $\eta_0$.

We focus on the diffusion coefficient $D$ and viscosity $\eta$.  In MD simulations, the diffusion coefficient $D$ is calculated using the Einstein relation from the mean square displacement (MSD) of a 15 ns NVT equilibrium trajectory at each pressure. The viscosity $\eta$ is calculated using the Green–Kubo  formalism, which relates $\eta$ to the integral of the stress autocorrelation function. Specifically, $\eta$ is given by
\beq
\begin{split}
\eta = \frac{1}{ Vk_{\rm B} T} \int_0^\infty \left\langle S^{\alpha \beta}(0) S^{\alpha \beta}(t) \right\rangle dt, \\
S^{\alpha \beta} = -\sum_i^{N} \left(m v_i^\alpha v_i^\beta + f_i^\alpha r_i^\beta \right),
\end{split}
\label{eq:GK}
\eeq
where \( S^{\alpha \beta}(t) \) (\( \alpha \neq \beta \)) denotes an off-diagonal component of the microscopic stress tensor, expressed as the sum of kinetic ($m v_i^\alpha v_i^\beta$) and virial ($f_i^\alpha r_i^\beta$) contributions. In our simulations, the stress autocorrelation function is computed from a longtime NVT trajectory of at least 15 ns. The time correlation is averaged over multiple time origins along the trajectory, ensuring the convergence of the Green–Kubo integral within the correlation time. For low-pressure systems, longer trajectories are used to capture the slower decay of stress fluctuations, and the average over multiple independent simulations with different initial conditions is performed to further reduce the statistical noise. For comparison, we also collect the viscosity data from the National Institute of Standards and Technology (NIST) database~\cite{NIST}, which are consistent with our simulation data within numerical accuracy.

To analyze the simulation results and NIST data (see Fig.~\ref{fig:eta}), we first examine the kinetic theory expressions, $D_{0} = \frac{3}{8\rho \sigma_0^2} \left(\frac{ k_{\rm B}T}{\pi m} \right)^{1/2}$
and
$\eta_0 = \frac{5}{16 \sigma_0^2} \left(\frac{mk_{\rm B}T}{\pi} \right)^{1/2}$, which are  valid in the dilute limit~\cite{chapman1990mathematical, hansen2013theory}. As shown in Fig.~\ref{fig:eta}(a), kinetic theory describes well the viscosity data below the $L^-$ line, where $\eta$ is a constant independent of pressure, as expected for gases. However, kinetic theory significantly fails above that line, as $\eta$ exhibits a rapid increase with pressure. The location where this happens is in very good agreement with the thermodynamic estimate of the $L^-$ crossover line \cite{li2024thermodynamic}.

To understand the origin of this deviation, one should note the approximations employed within such a formalism. Kinetic theory, as indicated by its name, only considers the kinetic effects, ignoring any explicit contributions from the interparticle potential.
The theory also neglects the memory effects in particle motions by assuming that successive collisions are completely uncorrelated. Mathematically, this means that the memory function of the stress autocorrelation function $G(t)$ is given by $M_G(t) \sim \delta(t)$, where $\delta(t)$ is the delta function (see Appendix~\ref{sec:theory}). As we explicitly demonstrate, both approximations fail above the $L^-$ line. Moreover, it is important to emphasize that the microscopic mechanisms underlying fluid viscosity differ fundamentally between gases and liquids. In gases, viscosity arises from particle collisions, whereas in liquids it is governed by potential energy barriers, processes that kinetic theory does not account for. This supports our proposal that the $L^-$ line marks the crossover where potential effects become significant, and the dynamics are no longer fully described by pair collisions, as in the gaseous regime.

In general, the viscosity contains two terms, $\eta = \eta_0 + \eta_{\rm ex}$, where $\eta_0$ is the kinetic contribution predicted by the kinetic theory, and $\eta_{\rm ex}$ comes purely from the interaction potential. Using this separation, the gas-SCF crossover line can be formally identified as the location at which $\eta_{\rm ex}\approx 0$ no longer holds. Lucas and Moser developed a formalism (hereafter referred to as LM theory) to compute $\eta_{\rm ex}$ based on a memory-function approach~\cite{lucas1979memory, lucas1979viscosity}. In order to achieve a more quantitative description, we take the version of the LM theory that approximates the excess memory function by a simple exponential form $M_G(t) = a e^{-bt}$, with $a$ and $b$ explicitly related to $g(r)$ and $V(r)$ (see Appendix~\ref{sec:theory}). 
As $P$ increases above $L^{-}$,  $\eta = \eta_0 + \eta_{\rm ex}$ quickly exceeds $\eta_0$. 
Using $g(r)$ from simulations, the theoretical values of $\eta$ computed using the LM theory are very close to simulation and NIST data [see Fig.~\ref{fig:eta}(a)]. Surprisingly, LM theory works reasonably well for both $\eta$ and $D$ even above the $L^+$ crossover lines. It remains unclear whether this is merely a fortuitous coincidence. 

We note that a related viscosity-based crossover definition has been proposed by Bell \textit{et al.}~\cite{Bell2021Dynamic}, who defined the gas-liquid demarcation as the locus where the kinetic and configurational contributions to viscosity are equal. Their criterion coincides with the minimum of the kinematic viscosity, i.e., the ratio of dynamic viscosity to density ($\nu = \eta / \rho$), which measures the fluid's momentum diffusivity and exhibits a characteristic minimum in the supercritical region due to competing density and viscosity effects. The crossover line identified by this criterion is close to the $L^+$ line, yet far from the $L^-$ line.

Among transport properties, the diffusion coefficient $D$ presents a more complex case since it is influenced by the structural order more indirectly. Because $D$ is related to the integral of the VACF, it does not have an excess term, in addition to the kinetic expression $D_{0} = \frac{3}{8\rho \sigma_0^2} \left(\frac{ k_{\rm B}T}{\pi m} \right)^{1/2}$. However, because this expression contains the equation of state (EOS) $\rho(P,T)$, which is connected to $g(r)$ and $V(r)$ through the virial equation, the kinetic expression $D_0$ depends on structure as well, despite in a more subtle way.

Indeed, $D_0$ computed using the full EOS agrees well with simulation data even beyond the $L^-$ line, extending approximately up to the $L^+$ line. In contrast, if the ideal gas equation of state is used, i.e., $\rho_{\rm ideal}(P,T) = P/(k_{\rm B} T)$, the resulting “ideal kinetic theory” begins to fail around the $L^-$ line [see Fig.~\ref{fig:eta}(b)]. This further supports our interpretation of $L^-$ as the boundary where structural correlations beyond nearest neighbors become significant, and the system departs from gas-like behavior. Moreover, this analysis highlights that the gas-SCF boundary can be identified through transport coefficients, whose behavior is influenced by the emergence of SSR structural order in the SCF phase.

\subsection{Liquid-SCF boundary: deviation from the Arrhenius law}

As already emphasized, above the liquid-SCF boundary (Frenkel-line), transport properties are dominated by a different microscopic mechanism. Meta-stable states in the potential landscape are formed, and the transitions between such states are realized by activation processes. In this regime, the Arrhenius law gives a relation between $D$ and the typical activation energy barrier $\Delta E$, $D \sim e^{-\Delta E/k_{\rm B}T} \sim e^{{\Delta S}/k_{\rm B}}$.   
The excess entropy ${\Delta S}$ can be approximately described by the two-particle entropy $S_2 = - 2\pi \rho k_{\rm B} \int_0^\infty \{g(r) \ln [g(r)] - [g(r)-1]\} r^2 dr$~\cite{Dzugutov1996Nature}, which can be directly computed from the $g(r)$ data. 
Leveraging these arguments, we find that the scaling $D \sim e^{{S_2}/k_{\rm B}}$ deviates from the simulation data around $L^+$, as expected [see Fig.~\ref{fig:eta}(c)]~\cite{demmel2025diffusion}.\\

\begin{figure}[h]
    \centering
  \includegraphics[width=0.96\linewidth]{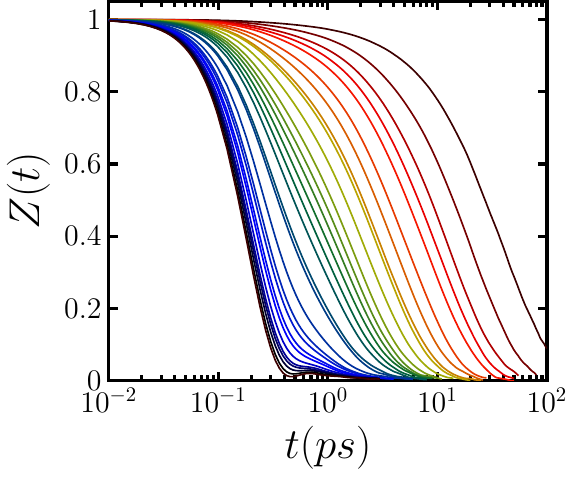}
    \caption{\textbf{Velocity autocorrelation functions.} Velocity autocorrelation functions $Z(t)$ for argon are obtained from MD simulations at $P=$\ 5,10,15,20,25, 35, 45, 55, 60, 75, 90, 100,110,120,135,150, 200,210,320,400,500,600,700,800,900,1000,1100,1200,1300\ \text{bar} (from right to left) and $T=195\ \text{K}$.
   }
    \label{fig:vacf}
\end{figure}

\begin{figure*}[t]
    \centering
    \includegraphics[width=\linewidth=0.9]{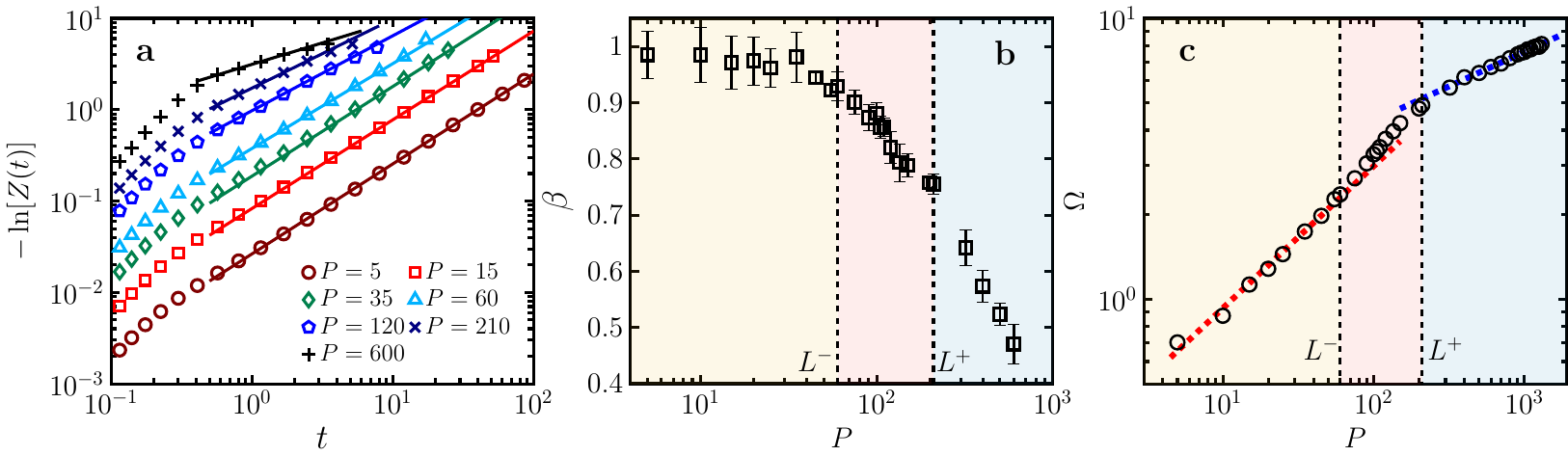}
    \caption{\textbf{Velocity autocorrelation functions.} \textbf{(a)} Fitting the tail of the VACF. Assuming a stretched exponential form, $Z(t) \sim e^{-(t/\tau)^\beta}$, the exponent $\beta$ is obtained by the slope of the linear fit in the log-log plot of $-\ln[Z(t)]$ vs. $t$.
    \textbf{(b)} $\beta$ as a function of $P$. Error bars represent the standard deviation of the fitted values from ten independent fitting ranges. \textbf{(c)} Einstein frequency $\Omega$ (THz) as a function of pressure $P$. The power-law fits (dashed colored lines) highlight the two distinct behaviors in the gas and liquid states.}
    \label{fig:VACF}
\end{figure*}

\section{\label{Dynamical fingerprints} Dynamical fingerprints: velocity autocorrelation function (VACF)}

\subsection{Gas-SCF boundary: crossover from exponential to stretched exponential decay}

Besides the signatures of the SCF state in structural and transport properties, the supercritical boundaries can also be identified by pure dynamic probes. 
We focus on the normalized VACF $Z(t)$ (Fig.~\ref{fig:vacf}), which satisfies $\dot{Z}(t) = - \int_0^t M_Z(t-\tau) Z(\tau) d\tau$, where $M_z(t)$ is the memory function. 

Here, we propose that the large-time behavior of the VACF, but also the very short one incarnated in the so-called Einstein frequency, contains information that reflects the gas-SCF boundary. It can be shown that (see Appendix~\ref{sec:theory}), when the memory function is $M_Z \sim \delta(t)$ (memory-less) or exponential $M_Z(t) \sim e^{-t/\tau_M}$, the VACF decays exponentially at large times $Z(t) \sim e^{-t/\tau}$. However, this simple exponential form is consistent with our simulation data only below the $L^-$ line.
Above the $L^-$ line, 
we find that at late time (up to the precision of our numerical data), the VACF is well-approximated by a stretched exponential,
    $Z(t)= A \exp \left[\left(- \frac{t}{\tau}\right)^\beta\right]$, with $\beta<1$.

In Fig.~\ref{fig:vacf}, we plot a complete set of VACFs measured by  MD simulations at $T=195$ K and different pressures. 
The VACF is fitted over a time window chosen to ensure convergence while minimizing numerical noise from the longtime tail. Because the convergence rate of the VACF strongly depends on pressure—being slower at low pressures and faster at high pressures—the fitting window for the stretchedexponential form is adjusted accordingly. For instance, at $5$~bar the VACF is evaluated over $100$~ps and fitted in ten consecutive $5$~ps segments starting from $5$~ps. As pressure increases and correlations decay more rapidly, progressively shorter fitting windows are used: $80$~ps at $10$~bar, $60$~ps at $15$~bar, $50$~ps at $20$--$25$~bar, and $4$--$6$~ps for pressures above $200$~bar. For the highest-pressure states ($\geq 400$~bar), the VACF converges within $4$~ps, and the fitting is performed in ten consecutive $0.1$~ps segments starting from $0.8$~ps.
The fitting is shown in Fig.~\ref{fig:VACF}(a).

 In Fig.~\ref{fig:VACF}(b) we show the  exponent $\beta$, obtained by fitting the data with the stretched exponential form, as a function of pressure. We find that the $L^-$ line coincides with the location at which $\beta$ approaches one. 
The stretched exponential of VACF (with $\beta<1$) suggests  
the onset of nontrivial memory effects beyond $L^-$, but its theoretical derivation deserves to be investigated in the future.

\subsection{Liquid-SCF boundary: disappearance of minimum in the velocity autocorrelations}

Interestingly, the intermediate-time behavior of the VACF has been used to 
locate the liquid-SCF boundary (the Frenkel-2 line). In fact, at high pressures, the VACF displays a minimum after a fast decay; this minimum is gradually suppressed by decreasing pressure, which eventually disappears around the crossover pressure defined by the Frenkel-2 line~\cite{brazhkin2013liquid}.

In addition, the signature of $L^\pm$ lines can also be found in the Einstein frequency $\Omega$ that appears in the very-short-time expansion of $Z(t) = 1-\Omega^2\frac{t^2}{2} + \cdots$ \cite{hansen2013theory} (see Fig.~\ref{fig:VACF}c). These results imply that, beyond structural and transport properties, the dynamical behavior of the SCF state also clearly distinguishes it from both the gas and liquid phases, reinforcing its identification as an intermediate
state of matter.\\

\begin{figure}[t]
    \centering
    \includegraphics[width=\linewidth]{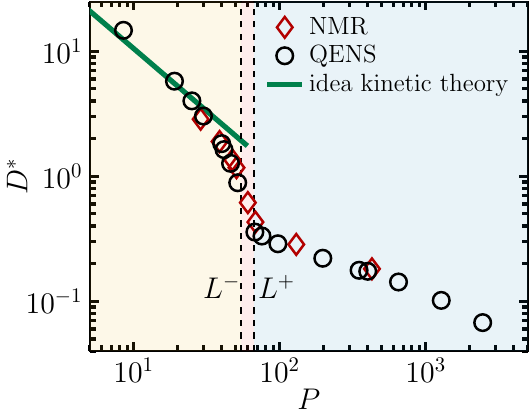}
    \caption{\textbf{Experimental diffusion coefficient for supercritical methane at $T=200$ K.}
    Experimental data obtained by quasielastic neutron scattering (QENS)~\cite{ranieri2024crossover} and nuclear magnetic resonance (NMR)~\cite{oosting1971proton} are shown, respectively, with open circles and open diamonds. The green line is the theoretical prediction from ideal kinetic theory, with scaling $D^* \propto P^{-1}$.}
    \label{fig:exp}
\end{figure}

\section{\label{Experimental signatures} Experimental signatures and proposals}
Our findings provide a basis for experimental verification in future studies. Accurate measurements of $g(r)$ in the SCF state are becoming available using modern neutron scattering instruments~\cite{dove2022pair, iwase2010newly} and the state-of-the-art data analysis tools~\cite{ma2020neudatool}. 

It is also possible to measure transport coefficients using experimental techniques such as neutron scattering. 
In fact, a recent experimental study has already reported accurate data of $D$ in supercritical methane along the $T=200$ K isotherm~\cite{ranieri2024crossover}. In Fig.~\ref{fig:exp}, we present the experimental data from Refs.~\cite{oosting1971proton,ranieri2024crossover}, together with the $L^{\pm}$ lines determined based on the NIST EOS of methane. The experimental data are fully consistent with our theoretical expectation. The diffusion constant $D$ follows the ideal kinetic theory behavior $D \sim 1/P$ up to a pressure around the $L^-$ line, as in Fig.~\ref{fig:eta}(b). Beyond the $L^-$ line a clear deviation from this scaling is observed. The SCF state between $L^\pm$ lines appears to coincide with a ``transition regime'' between the gaseous low-$P$ and the liquid high-$P$ behavior. In agreement with our simulation results, this observation provides experimental evidence of the two $L^\pm$ crossover lines and of the existence of the supercritical fluid state, clearly separated from the gas and liquid phases. During the finalization of our work, Ref.~\cite{kpsb-wvxl} appeared, presenting a parallel theoretical analysis of the experimental data from Ref.~\cite{ranieri2024crossover}. At present, any connection between our results and the ideas put forward in Ref.~\cite{kpsb-wvxl} remains unclear, but certainly warrants further investigation.

Finally, the signatures of the supercritical boundary lines could be experimentally observed in a more phenomenological way. Enhanced density fluctuations have been observed in small-angle neutron scattering experiments around the $L^+$ line~\cite{li2024thermodynamic} in supercritical carbon dioxide~\cite{pipich2018densification}, due to the formation of liquid-like droplets. Analogously, we also expect enhanced density fluctuations around the $L^-$ line, due to the formation of gas-like bubbles. If one interprets the supercritical transition regime between $L^\pm$ lines as the blur-up of the subcritical discontinuous first-order transition, then the rapid change of the EOS in this regime might be related to the phenomenon of supercritical pseudoboiling~\cite{banuti2015crossing, wang2021three, he2023distinguishing}.\\

\section{Comparison to other scenarios}
\subsection{Comparison to the Widom line scenario: a single separation line between liquid-like and gas-like states }

The Widom line is defined as the locus of maximum correlation length~\cite{xu2005relation}, which can be extracted from the static structure factor. In order to compare $L\pm$ with the Widom line, we compute the static structure factor $S(q)$ directly from MD trajectories. For each configuration snapshot containing $N = 4000$ argon atoms, $S(\mathbf{q})$ is calculated as

\begin{equation}
    S(\mathbf{q}) = \frac{1}{N} \left| \sum_{j=1}^{N} e^{-i\mathbf{q}\cdot\mathbf{r}_j} \right|^2,
\label{eq:sq_cal}
\end{equation}
where $\mathbf{r}_j$ are atomic positions and $\mathbf{q}$ are wave vectors compatible with the periodic boundary conditions. The wave vectors are grouped into spherical shells of width $\Delta q = 0.05$~\AA$^{-1}$, and $S(\mathbf{q})$ values within each shell are averaged to obtain $S(q)$. For each pressure, $S(q)$ is evaluated over 500 ps and averaged over 500\,000 configurations.

The Ornstein-Zernike (OZ) correlation length is obtained by fitting the low-$q$ region of $S(q)$ to the OZ form:
\begin{equation}
    S(q) = \frac{S_0}{1 + \xi_{\mathrm{OZ}}^2 q^2}.
\label{eq:Ozform}
\end{equation}
Inverting this relation, one gets
\begin{equation}
    \frac{1}{S(q)} = \frac{1}{S_0} + \frac{\xi_{\mathrm{OZ}}^2}{S_0} q^2,
\label{eq:Ozlinearform}
\end{equation}
from which $\xi_{\mathrm{OZ}}$ is determined via linear regression. Uncertainties are estimated from the fitting errors.

\begin{figure}[ht]
    \centering
    \includegraphics[width=\linewidth]{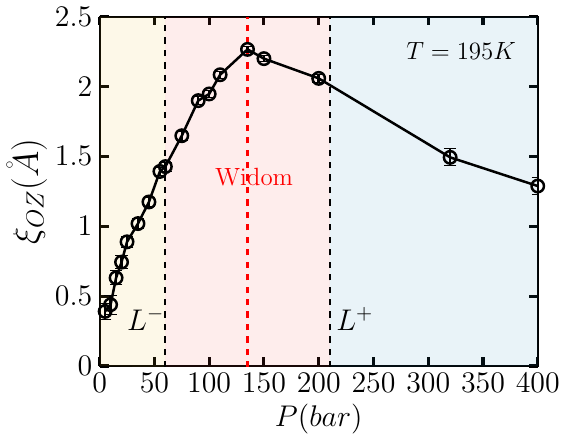}
    \caption{\textbf{Ornstein-Zernike correlation length.} 
    $\xi_{\mathrm{OZ}}$ as a function of pressure at $T = 195$~K, showing a maximum near the Widom line between the $L^+$ and $L^-$ lines. Error bars represent fitting uncertainties.}
    \label{fig:oz}
\end{figure}

The resulting $\xi_{\mathrm{OZ}}$ as a function of pressure is shown in Fig.~\ref{fig:oz}. $\xi_{\mathrm{OZ}}$ reaches its maximum near the Widom line, which is located between the $L^\pm$ lines.

\subsection{Comparison to the Widom delta scenario: a coexistence region of liquid-like and gas-like atoms}

\begin{figure*}
    \centering
    \includegraphics[width=0.8\linewidth]{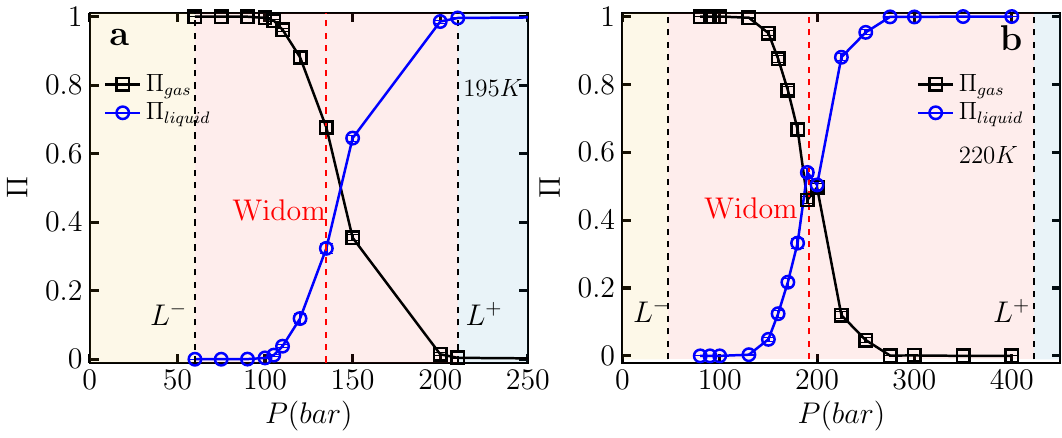}
    \caption{\textbf{Comparison between Widom delta and $L^\pm$ lines.} Fractions of gas-like ($\Pi_{\text{gas}}$) and liquid-like ($\Pi_{\text{liquid}}$) atoms as functions of pressure for simulated argon at two temperatures: $\mathbf{(a)}$ $T=195\,\mathrm{K}$ and $\mathbf{(b)}$ $T=220\,\mathrm{K}$. The red dashed line indicates the Widom line.
}
    \label{fig:perc}
\end{figure*}

We also compare $L^{\pm}$ with the Widom delta. In this framework, however, the intermediate SCF region is not treated as a distinct state of matter but rather as a gas–liquid ``coexistence'' regime. Following the approach of Yoon \textit{et al.} \cite{2018Probabilistic}, we employ a probabilistic classification algorithm to characterize the gas-like and liquid-like domains in the simulated argon systems. 

For each configuration from MD data, we perform the Voronoi tessellation. The Voronoi cell of atom $i$ is defined as the set of points closer to atom $i$ than to any other atom. The local density $\rho_i^{\text{local}}$ is then calculated as the inverse of the Voronoi cell volume,
\begin{equation}
    \rho_i^{\text{local}} = \frac{1}{V_i^{\text{Voronoi}}},
\end{equation}
where $V_i^{\text{Voronoi}}$ is the volume of the Voronoi polyhedron associated with atom $i$.

To account for spatial correlations and reduce sensitivity to local fluctuations, we compute the mean local density $\bar{\rho}_i$ for each atom $i$ by averaging over its Voronoi neighbors,
\begin{equation}
    \bar{\rho}_i = \frac{1}{N_{\text{neigh}} + 1} \left( \rho_i^{\text{local}} + \sum_{j \in \mathcal{N}(i)} \rho_j^{\text{local}} \right),
\end{equation}
where $\mathcal{N}(i)$ denotes the set of Voronoi neighbors of atom $i$ (atoms sharing a Voronoi face), and $N_{\text{neigh}} = |\mathcal{N}(i)|$ is the number of such neighbors.

For each pressure, we calculate the fractions of gas-like and liquid-like atoms as ensemble averages (see Appendix~\ref{sec:Widom_delta}). The gas-like fraction $\Pi_{\text{gas}}$ is defined as
\begin{equation}
    \Pi_{\text{gas}} = \frac{1}{N_{\text{atoms}}} \sum_{i=1}^{N_{\text{atoms}}} \mathcal{I}\left(P_i^{\text{liq}} \leq 0.5\right),
    \label{eq:pi_gas}
\end{equation}
where $\mathcal{I}(\cdot)$ is the indicator function. Similarly, the liquid-like fraction $\Pi_{\text{liq}}$ is
\begin{equation}
    \Pi_{\text{liq}} = 1 - \Pi_{\text{gas}}.
    \label{eq:pi_liq}
\end{equation}
The fractions are averaged over 100 random configurations along the trajectories. Uncertainties in these fractions are estimated from the standard deviation across multiple simulation frames.

The resulting fractions of gas-like and liquid-like atoms as functions of pressure for simulated argon at two temperatures are shown in Fig.~\ref{fig:perc}. 
The Widom delta is defined from the inverse sigmoidal fit to $\Pi_{\text{gas}}$ by finding the densities (pressures) where the tangent line at $\Pi_{\text{gas}} = 0.5$ intersects $\Pi_{\text{gas}} = 1.0$ and $\Pi_{\text{gas}} = 0.0$, respectively. We find that the Widom delta is within the intermediate transition regime bounded by $L^\pm$ lines.

\section{\label{Discussion} Discussion}
In this study, we demonstrate that the signatures of the supercritical fluid boundaries can be identified by various physical quantities, including structure descriptors such as the radial distribution function, transport coefficients, and the velocity autocorrelation function. 
Our study suggests
the phase diagram above the critical point, clarifying the existence of three independent states of matter characterized by distinct physical properties. (i) The liquid-SCF boundary is the limit beyond which the system does not behave like a typical liquid.
Because the liquid state can be characterized by various features, ranging from thermodynamics to dynamics to structures, the deviation from the liquid behavior can be defined according to different criteria~\cite{fisher1969decay, PhysRevE.85.031203, brazhkin2013liquid, PhysRevE.85.031203}. (ii) The gas-SCF boundary is the limit beyond which the system does not behave like a typical gas. This line is reflected in several physical properties and can therefore be defined using thermodynamics, structural order, transport coefficients, and finally dynamical behavior. (iii) In-between is the SCF state that is neither liquid-like nor gas-like but rather a separate state of matter, whose properties cannot be easily rationalized as a simple linear superposition of gas and liquid features. We find that the Widom line defined by the maximum of the Ornstein-Zernike correlation length~\cite{xu2005relation}, and the Widom delta identified according to the fractions of liquid-like and gas-like atoms~\cite{2018Probabilistic}, both lie within the intermediate region bounded by $L^\pm$ lines.

It remains unclear whether our findings, and in particular the existence of a gas-SCF boundary, extend to systems with purely repulsive interactions, where no thermodynamic separation between liquid and gas phases, and no critical point, exists. So far, in these systems, only a boundary line between gas-like and liquid-like states has been discussed, e.g., \cite{huang2023revealing,Jiang2025}. Moreover, whether our results extend to more complex molecular fluids such as CO$_2$ or H$_2$O remains an open question and warrants further investigation.

On the other hand, the proposed framework and analysis methods can be generalized to many other systems. In particular, it would be very interesting to study the nature of order in the quantum supercritical state~\cite{wang2024quantum, lv2024quantum}, in the context of quantum many-body correlations. Moreover, the dynamical signatures of the $L^-$ line can be explored  in the supercritical kinetics of black holes~\cite{Zhao:2025ecg,li2025critical} and holographic models as well \cite{Zhao:2024jhs}. Another future direction is to examine in more detail the analogy between the supercritical fluid state and the QCD phase diagram of nuclear matter. Interestingly, a three-state phase diagram presenting an intermediate partially deconfined phase, and naive similarities with our proposal depicted in Fig.~\ref{fig:1}, has been recently discussed \cite{Hanada2023,Glozman2022,fujimoto2025new}.   

In conclusion, returning to the fundamental question of ``what separates a liquid from a gas''~\cite{10.1063/PT.3.1796}, our revised answer points to the existence of an intermediate supercritical fluid state, distinguishable from both gas and liquid phases in terms of thermodynamic, structural, transport, and dynamical properties.

\begin{acknowledgments}
We thank S. Khrapak for bringing the recent Ref.~\cite{kpsb-wvxl} to our attention. S.J. and M.B. acknowledge the support of the Shanghai Municipal Science and Technology Major Project (Grant No. 2019SHZDZX01). X.L. acknowledges the China Postdoctoral Science Foundation (Grant No. 2026M794653), the support of the National Natural Science Foundation of China (Grant No. 12605057) and the Postdoctoral Fellowship Program of CPSF (Grant No. GZC20252776). X.F. acknowledges the support of the National Natural Science Foundation of China (Grant No. 52201016). M.B. acknowledges the support of the Yangyang Development Fund. Y.J. acknowledges funding from the National Key R\&D Program of China (Grant No. 2025YFF0512000), the project supported by the Space Application System of China Manned Space Program, NSFC (Grant No. 12447101), and the Wenzhou Institute (Grant No. WIUCASICTP2022). S.J. and X.F. thank the Key State Laboratory for Clean and Efficient Turbomachinery Power Equipment for providing High Performance Cluster resources. The authors acknowledge the use of the High Performance Cluster at the Institute of Theoretical Physics, Chinese Academy of Sciences, and the computer clusters at the Hefei Advanced Computing Center.
\end{acknowledgments}

{\bf Data availability.}
There are no publicly available research data or software supporting this manuscript. Requests for further information or data should be sent to the authors.

\appendix

\section{Theories of autocorrelation functions and transport coefficients}
\label{sec:theory}

\subsection{Velocity autocorrelation function (VACF) and diffusion coefficient} The (normalized) VACF is defined by
\beq
Z(t) \equiv \frac{\left \langle  {\bf v}(0) \cdot  {\bf v}(t) \right \rangle}{\left \langle{\bf v}(0) \cdot {\bf v}(0) \right \rangle}, 
\eeq
and the diffusion coefficient $D$ is related to the VACF through the Green-Kubo formula,
\beq
D = \frac{k_{\rm B}T}{m}\int_0^\infty Z(t) dt, 
\label{eq:D_def}
\eeq
where $T$ is the temperature, $m$ the mass of particle, $k_{\rm B}$ the Boltzmann constant.
In general, the VACF satisfies the following equation,
\beq
\dot{Z}(t) = - \int_0^t M_Z(t-\tau) Z(\tau) d\tau,
\label{eq:Z_eq}
\eeq
where $M_Z(t)$ is the memory function describing the memory effects in the motion of particles, and $\dot{Z}$ represents the time derivative of $Z$. 

\subsubsection{Kinetic theory}
In the dilute limit, the dynamics are memory-less, 
\beq
M_{Z}(t) = \frac{1}{\tau} \delta(t),
\label{eq:M_delta}
\eeq
 and the VACF decays exponentially, 
\beq
Z(t) =  e^{-t/\tau},
\eeq
where $\tau$ is the velocity correlation time. 
Equation~(\ref{eq:D_def}) then gives
\beq
D = \frac{k_{\rm B}T \tau}{m}.
\eeq
Elementary kinetic theory~\cite{chapman1990mathematical} shows that in the dilute limit, $\tau = l/\langle v \rangle$, where $l$ is the mean free path and $\langle v \rangle$ the average velocity. For hard spheres, the kinetic theory  gives,
$\tau = \frac{3}{8\rho \sigma_0^2} \left(\frac{\pi k_{\rm B}T}{m} \right)^{-1/2}$,
where $\sigma_0$ is the diameter of hard spheres, and
\beq
D_{0} = \frac{3}{8\rho \sigma_0^2} \left(\frac{ k_{\rm B}T}{\pi m} \right)^{1/2}.
\label{eq:D0}
\eeq
For the Lennard-Jones potential, $r= \sigma_0 \approx 1.12 \sigma$  is where the potential energy $V(r)$ (see Eq.~\ref{eq:V_LJ})  minimizes.

\subsubsection{Lucas-Moser theory}
At high densities, the memory-less assumption, Eq.~\eqref{eq:M_delta}, does not hold anymore. However, in the interested regime, i.e., for the SCF around the gas-SCF boundary, the memory effects are not significant, and the memory function $M_Z(t)$ decays much faster than the VACF $Z(t)$. In this case, one can assume a simple form of $M_Z(t)$, such as the following exponential form:
\beq
M_Z = \Omega^2 e^{-t/\tau_M}.
\label{eq:MZ}
\eeq
Here, $\Omega$ is the Einstein frequency that appears in the short-time expansion of $Z(t) = 1-\Omega^2\frac{t^2}{2} + \cdots $, and $\tau_M$ is the characteristic time of the memory function. Plugging Eq.~(\ref{eq:MZ}) into Eq.~(\ref{eq:Z_eq}), one obtains
\beq
Z(t) = \frac{1}{\alpha_+ - \alpha_-}\left(\alpha_+ e^{-\alpha_-t} - \alpha_- e^{-\alpha_+ t}\right),
\label{eq:Z_LM}
\eeq
where
\beq
\alpha_{\pm} = \frac{1}{2\tau_M} \left[1 \mp \left(1 - 4 \Omega^2 \tau_M^2 \right)^{1/2} \right].
\eeq
Equation~(\ref{eq:D_def}) then gives
\beq
D = \frac{k_{\rm B}T}{m \Omega^2 \tau_M}.
\label{eq:D_tauM}
\eeq
In order to explicitly compute $D$, one needs to know $\Omega^2$ and $\tau_M$. Lucas and Moser have related these two quantities to  $g(r)$ and the interparticle potential $V(r)$, through the following formulas~\cite{lucas1979memory}:
\beq
\Omega^2 = \frac{\left \langle \dot{\bf v}(0) \cdot \dot{\bf v}(0) \right \rangle}{\left \langle {\bf v}(0) \cdot {\bf v}(0) \right \rangle},
\eeq
\beq
\frac{1}{\tau_M} = \left [ \frac{\left \langle \ddot{\bf v}(0) \cdot \ddot{\bf v}(0) \right \rangle}{\left \langle \dot{\bf v}(0) \cdot \dot{\bf v}(0) \right \rangle} - \frac{\left \langle \dot{\bf v}(0) \cdot \dot{\bf v}(0) \right \rangle}{\left \langle {\bf v}(0) \cdot {\bf v}(0) \right \rangle} \right]^{1/2},
\eeq

\begin{widetext}
where
\begin{equation}
\left \langle {\bf v}(0) \cdot {\bf v}(0) \right \rangle = \frac{3k_{\rm B}T}{m},
\end{equation}

\begin{equation}
\frac{\left \langle \dot{\bf v}(0) \cdot \dot{\bf v}(0) \right \rangle}{3k_{\rm B}T/m\sigma_0} 
=32 \pi \frac{n^*}{T^*} \int_0^{\infty} g_2\left(r_{12}^*\right)\left[3 W_{12}^{\prime}+2 W_{12}^{\prime \prime} r_{12}^{* 2}\right] r_{12}^{* 2} d r_{12}^*,
 \end{equation}
\begin{equation}
\begin{aligned}
&\frac{\left \langle \ddot{\bf v}(0) \cdot \ddot{\bf v}(0) \right \rangle}{3(k_{\rm B}T)^2(\epsilon/m^3\sigma_0^4)}
= 512 \pi \frac{n^*}{T^*} \int_0^{\infty} g_2\left(r_{12}^*\right)\left[3 W_{12}^{\prime} +4 W_{12}^{\prime} W_{12}^{\prime \prime} r_{12}^{* 2}+4 W_{12}^{\prime \prime} r_{12}^{* 4}\right] r_{12}^{* 2} d r_{12}^*\\
\label{eq:ddv2}
\end{aligned}
 \end{equation}

with 
\beq
\begin{aligned}
W_{i j}  =\frac{V(r)}{4 \epsilon} ,\quad
W_{i j}^{\prime}  =d W_{i j} / d r_{i j}{ }^{* 2},\quad
W_{i j}^{\prime \prime}  =d^2 W_{i j} /\left(d r_{i j}{ }^{* 2}\right)^2,\quad 
n^*  =n \sigma_0^3,\quad
T^*  =\frac{k_{\rm B} T}{\epsilon}.
\end{aligned}
\eeq
\end{widetext}
Equations~(\ref{eq:D_tauM})-(\ref{eq:ddv2}) give a theoretical estimate of the diffusion coefficient, $D_{\rm LM}$. Compared to the kinetic theory $D_0$,  $D_{\rm LM}$ takes into account the dynamical memory effects [Eq.~(\ref{eq:MZ})] and the spatial pair correlations $g(r)$.

Unsurprisingly, even the Lucas-Moser theory breaks down if the density is further increased.  The VACF Eq.~(\ref{eq:Z_LM}) still has an exponential form at large times, with a correlation time $\tau \sim 1/\alpha_-$ (note that $\alpha_+ < \alpha_-$ ). As shown in the main text, this exponential form does not hold anymore at high densities---instead, we need a stretched exponential function to fit the simulation data. However, the exponential approximation does not seem to affect the transport coefficients significantly. 

\subsection{Stress autocorrelation function and viscosity} 
The viscosity is connected to the stress autocorrelation function,
\beq
G(t) \equiv \left \langle  S^{\alpha \beta}(0) S^{\alpha \beta}(t) \right \rangle,
\label{eq:G_def}
\eeq
via the Green-Kubo relation,
\beq
\eta = \frac{1}{V k_{\rm B}T} \int_0^\infty G(t) dt.
\eeq
Here $S^{\alpha \beta}$ represents the $\alpha \beta$ component of the off-diagonal ($\alpha \neq \beta$) stress tensor, defined as
\beq
S^{\alpha \beta} = - \sum_i^{N} \left(m v_i^\alpha v_i^\beta + f_i^\alpha r_i^\beta \right) = S^{\alpha \beta}_0 + S_{\rm ex}^{\alpha \beta},
\label{eq:S_def}
\eeq
where $f_i^\alpha$ is the net force of all particles on the $i$-th particle in the $\alpha$-direction, and $r_i^\beta$ is the coordinate of the  
$i$-th particle in the $\beta$-direction. For the convenience of discussion below, we separate the kinetic stress $S_0^{\alpha \beta}$ that depends on velocities, and the excess stress $ S_{\rm ex}^{\alpha \beta}$ that depends on the forces (the potential energy). Because the considered fluid is isotropic, we assume $S^{xy} = S^{yz} = S^{xz}$. 
\subsubsection{Kinetic theory} 
Using Eqs.~(\ref{eq:G_def}) and~(\ref{eq:S_def}), we can expand $G(t)$ into four terms,
\beq
\begin{split}
G(t)  & =  \left \langle  S_0^{\alpha \beta}(0) S_0^{\alpha \beta}(t) \right \rangle
+ \left \langle S_{\rm ex}^{\alpha \beta}(0)S_{\rm ex}^{\alpha \beta}(t) \right \rangle \\
&+ \left \langle  S_0^{\alpha \beta}(0)S_{\rm ex}^{\alpha \beta}(t) \right \rangle
+ \left \langle S_{\rm ex}^{\alpha \beta}(0) S_0^{\alpha \beta}(t) \right \rangle \\
& \approx \left \langle  S_0^{\alpha \beta}(0) S_0^{\alpha \beta}(t) \right \rangle
+ \left \langle S_{\rm ex}^{\alpha \beta}(0)S_{\rm ex}^{\alpha \beta}(t) \right \rangle \\
& \equiv G_0(t) +G_{\rm ex}(t).
\end{split}
\eeq

The third and fourth terms in the first line are neglected, assuming that the kinetic and potential parts are uncorrelated. In the kinetic theory, the excess part $G_{\rm ex}(t)$ is also ignored. In this case, $\eta$ is determined solely by velocity autocorrelations. Under similar approximations (e.g., the dynamics are memoryless), one can derive~\cite{chapman1990mathematical},
\beq
\eta_0 = \frac{5}{16\rho \sigma_0^2} \left(\frac{ mk_{\rm B}T}{\pi} \right)^{1/2},
\label{eq:eta0}
\eeq
which has a simple relation with $D_0$, $\eta_0 \sim D_0 \rho$.

\subsubsection{Lucas-Moser theory}
Lucas and Moser have also computed the excess viscosity by considering an exponential memory function~\cite{lucas1979viscosity}. For the excess part, the stress autocorrelation function is
\beq
G_{\rm ex}(t) \equiv
\left \langle S_{\rm ex}^{\alpha \beta}(0)S_{\rm ex}^{\alpha \beta}(t) \right \rangle,
\eeq
which satisfies
\beq
\dot{G}_{\rm ex}(t) = - \int_0^t M_G(t-\tau)G_{\rm ex}(\tau) d\tau.
\label{eq:G_eq}
\eeq
Using an exponential ansatz,  
\beq
M_G(t)= a e^{-b t},
\eeq
where 
\beq
a = \frac{\left \langle \dot{S}^{\alpha\beta}(0)  \dot{S}^{\alpha\beta}(0) \right \rangle}{\left \langle {S}^{\alpha\beta}(0)  {S}^{\alpha\beta}(0) \right \rangle},
\eeq

\beq
b = \left [ \frac{\left \langle \ddot{S}^{\alpha\beta}(0)  \ddot{S}^{\alpha\beta}(0) \right \rangle}{\left \langle \dot{S}^{\alpha\beta}(0)  \dot{S}^{\alpha\beta}(0) \right \rangle} - \frac{\left \langle \dot{S}^{\alpha\beta}(0)  \dot{S}^{\alpha\beta}(0) \right \rangle}{\left \langle {S}^{\alpha\beta}(0)  {S}^{\alpha\beta}(0) \right \rangle} \right ]^{1/2},
\eeq
the excess viscosity can be evaluated, 
\beq
\eta_{\rm ex} = \frac{b}{Vk_{\rm B}Ta},
\eeq
and the total viscosity is,
\beq
\eta = \eta_0 + \eta_{\rm ex}.
\eeq
Note that the $\eta_{\rm ex}$ depends on $g(r)$ through the following expressions: 
\begin{widetext}
\beq
\frac{\left \langle {S}^{\alpha\beta}(0)  {S}^{\alpha\beta}(0) \right \rangle}{V(k T)^2\left(1 / \sigma_0^3\right)} = n^*+\frac{16 \pi}{15} \frac{n^{* 2}}{T^*} \int_0^{\infty} g_2\left(r_{12}^*\right)\left[5 W_{12}^{\prime}+2 W_{12}^{\prime \prime} r_{12}^{* 2}\right] r_{12}^{* 4} d r_{12}^*,
\eeq
\beq
\begin{aligned}
&\frac{\left \langle \dot{S}^{\alpha\beta}(0)  \dot{S}^{\alpha\beta}(0) \right \rangle}{V(k T)^2\left(\epsilon / m \sigma_0^5\right)}  =\frac{1024 \pi}{15} \frac{n^{* 2}}{T^*} \int_0^{\infty} g_2\left(r_{12}^*\right)\left[10 W_{12}^{\prime} +4 W_{12}^{\prime} W_{12}^{\prime \prime} r_{12}^{* 2} +W_{12}^{\prime \prime} r_{12}^{* 4}\right] r_{12}^{* 4} d r_{12}^*,
\end{aligned}
\eeq
\beq
\begin{aligned}
\frac{\left \langle \ddot{S}^{\alpha\beta}(0)  \ddot{S}^{\alpha\beta}(0) \right \rangle}{V (k T)^2\left(\epsilon^2 / m^2 \sigma_0^7\right)} =&\frac{2048 \pi}{15} \frac{n^{* 2}}{T^*} \int_0^{\infty} g_2\left(r_{12}{ }^*\right)\left[T ^ { * } \left(90 W_{12}^{\prime}+180 W_{12}^{\prime} W_{12}^{\prime \prime} r_{12}{ }^{* 2}\right.\right.+24 W_{12}^{\prime} W^{\prime \prime \prime}{ }_{12} r_{12}{ }^{* 4} \\
&+231 W^{\prime \prime}{ }_{12}{ }^2 r_{12}{ }^{* 4}+ 84 W^{\prime \prime}{ }_{12} W^{\prime \prime \prime}{ }_{12} r_{12}{ }^{* 6} \left.+12 W^{\prime \prime \prime}{ }_{12}{ }^2 r_{12}{ }^{* 8}\right)+4\left(20 W_{12}^{\prime}+24 W_{12}^{\prime} W^{\prime \prime}{ }_{12} r_{12}{ }^{* 2}\right.\\
&\left.\left.+18 W_{12}^{\prime} W^{\prime \prime}{ }_{12}{ }^2 r_{12}{ }^{* 4}+4 W^{\prime \prime}{ }_{12}{ }^3 r_{12}{ }^{* 6}\right) r_{12}{ }^{* 2}\right] r_{12}{ }^{* 2} d r_{12}{ }^* ,
\end{aligned}
\eeq

\end{widetext}
where
\beq
\begin{aligned}
W_{i j}^{\prime \prime \prime} & =d^3 W_{i j} /\left(d r_{i j}{ }^{* 2}\right)^3.
\end{aligned}
\eeq

\section{Classification of gas-like and liquid-like atoms}
\label{sec:Widom_delta}

The classification of atoms into gas-like and liquid-like states follows a probabilistic algorithm inspired by the work of Yoon \textit{et al.} \cite{2018Probabilistic}:

\begin{enumerate}
    \item \textbf{Random processing order.} Atoms are processed in random order. When an atom is selected, the mean local density of that atom and all its Voronoi neighbors is computed. This neighborhood averaging reduces sensitivity to local fluctuations.
    
    \item \textbf{Threshold comparison.} For the current unlabeled atom $i$, its mean local density $\bar{\rho}_i$ is compared to a threshold value $\rho_{\text{th}}$. We use the critical density of argon $\rho_c$ as the threshold,
    \begin{equation}
        \text{label} = 
        \begin{cases}
            1 & (\text{liquid-like}) \quad \text{if } \bar{\rho}_i \geq \rho_c, \\
            0 & (\text{gas-like}) \quad \text{if } \bar{\rho}_i < \rho_c.
        \end{cases}
    \end{equation}
    
    \item \textbf{Simultaneous labeling.} The chosen atom $i$ and all its unlabeled Voronoi neighbors are assigned the same label, reflecting the correlated nature of local environments.
    
    \item \textbf{Multiple trials.} Steps 1--3 are repeated $N_{\text{trials}} = 300$ times with different random processing orders to obtain statistically robust results.
    
    \item \textbf{Probability calculation.} The liquid-like probability $P_i^{\text{liq}}$ for atom $i$ is calculated as the fraction of trials in which it is classified as liquid-like,
    \begin{equation}
        P_i^{\text{liq}} = \frac{1}{N_{\text{trials}}} \sum_{t=1}^{N_{\text{trials}}} L_i^{(t)},
    \end{equation}
    where $L_i^{(t)} = 1$ if atom $i$ is labeled liquid-like in trial $t$, and $L_i^{(t)} = 0$ otherwise.
\end{enumerate}

\bibliography{biblio}

\end{document}